# Efficient and Secure Routing Protocol for Wireless Sensor Networks through SNR based Dynamic Clustering Mechanisms

*S.Ganesh[1], R.Amutha[2]

[1]Sathyabama University, Chennai, Tamil Nadu, India

[2]SSN College of Engineering, Chennai, Tamil Nadu, India.

[1]ganesh8461@gmail.com

[2]amuthar@ssn.edu.in

**Abstract**: Advances in Wireless Sensor Network Technology (WSN) have provided the availability of small and low-cost sensor with capability of sensing various types of physical and environmental conditions, data processing and wireless communication. In WSN, the sensor nodes have a limited transmission range, and their processing and storage capabilities as well as their energy resources are limited. Triple Umpiring System (TUS) has already been proved its better performance on Wireless Sensor Networks. Clustering technique provides an effective way to prolong the lifetime of WSN. In this paper, we modified the Ad hoc on demand Distance Vector Routing (AODV) by incorporating **Signal to Noise Ratio (SNR)** based dynamic clustering. The proposed scheme Efficient and Secure Routing Protocol for Wireless Sensor Networks through SNR based dynamic Clustering mechanisms (ESRPSDC) can partition the nodes into clusters and select the Cluster Head (CH) among the nodes based on the energy and Non Cluster Head (NCH) nodes join with a specific CH based on SNR Values. Error recovery has been implemented during Inter cluster routing itself in order to avoid end-to-end error recovery. Security has been achieved by isolating the malicious nodes using sink based routing pattern analysis. Extensive investigation studies using Global Mobile Simulator (GloMoSim) showed that this Hybrid ESRP significantly improves the Energy efficiency and Packet Reception Rate (PRR) compared to SNR unaware routing algorithms like Low Energy Aware Adaptive Clustering Hierarchy (LEACH) and Power-Efficient Gathering in Sensor Information Systems (PEGASIS).

*Author for correspondence

## I. Introduction

Sensor Network Wireless is widely considered as one of the most important technologies for the twenty-first century. The sensing electronics measure ambient conditions related to the environment surrounding the sensors and transform them in to an electrical signal. In many WSN applications, the deployment of sensor nodes is performed in an ad-hoc fashion without careful planning and engineering. In the past few years, an intensive research that addresses the potential of collaboration among sensors in data gathering and processing and in the coordination and management of the sensing activities were conducted. However, sensor nodes are constrained in energy supply and bandwidth.

Energy conservation is critical in Wireless Sensor Networks. Replacing or recharging batteries is not an option for sensors deployed in hostile environments. Generally, communication electronics in the sensor utilizes most energy. Stability is one of the major concerns in advancement of Wireless Sensor Networks (WSN). A number of applications of WSN require guaranteed sensing, coverage and connectivity throughout its operational period. Death of the first node might cause instability in the network. Therefore, all of the sensor nodes in the network must be alive to achieve the goal during that period. One of the major obstacles to ensure these phenomena is unbalanced energy consumption rate. Numerous techniques were proposed to improve energy consumption rate such as clustering, efficient routing, and data aggregation.

In a typical WSN application, sensor nodes are scattered in a region from where they collect data to achieve certain goals. Data collection may be continuous, periodic or event based process. WSN must be very stable in some of its applications like security monitoring and motion tracking.

Death of only one sensor node may disrupt coverage or connectivity and thus may reduce stability in this sort of applications. Therefore, all of the deployed sensor nodes in WSN must be active during operational lifetime. However, sensor nodes are generally equipped with one-time batteries and most of the batteries are of low energy. For this reason, each sensor node must efficiently use its available energy in order to improve the lifetime of WSN. Different techniques are used for efficient usage of this low available energy in a sensor node. Clustering is one of these most well known techniques.

Li et al [1] have investigated the joint Power Allocation (PA) issue in a class of MIMO relay systems. By using the capacity and the mean-square error (MSE) as optimization criterion, two joint PA optimization problems have been formulated. As the cost functions derived directly from the capacity and the MSE would lead to nonconvex optimization, two modified cost functions corresponding to a convex problem of the source and the relay power weighting coefficients have been developed. The key contribution of the proposed method lies in the discovery of a tight bound for the capacity and the MSE that simplifies the joint source and relay power allocation into a convex problem. A distinct feature of the new method is that the power allocation within the source and that within the relay are jointly optimal for any given power ratio of the two units.

It was studied in [2] that the joint power allocation problem for multicast systems can achieve better data rate. To deal with the nonconvex optimization problem, a high-SNR approximation is employed to modify the original cost function in order to obtain a convex minimization problem, where the approximation is shown to be asymptotically optimal at the high-SNR regime. As an alternative, an iterative algorithm has been developed by utilizing the convexity property of the cost function with respect to a part of the whole power coefficients. Considering the low complexity of the physical layer network coding in the multi-cast system, the lattice based network coding that uses the proposed joint power allocation schemes has been suggested.

In this paper, we have developed a Hybrid Efficient and Secure Routing Protocol through SNR based dynamic Clustering mechanisms (ESRPSDC), which is a combination of SNR, based dynamic Clustering and routing pattern based security mechanisms. We did a brief comparison of ESRPSDC with LEACH and PEGASIS, two of the popular routing protocols. The rest of this paper is organized as follows In Section 2, the related work is briefly reviewed and discussed. Then we describe our network model, adversary model and notations used throughout in this paper in Section 3. Simulation Results are presented in Section 4. We conclude this paper in Section 5.

## II. Related Work

Several techniques have already been proposed to improve network lifetime in WSN. Among them, clustering in one of the widely accepted techniques. Clustering is also used in wireless adhoc networks, mobile adhoc networks along with sensor networks. Several clustering techniques have already been introduced for partitioning nodes in to areas.

Clustering is a technique in which deployed sensor nodes are grouped into some clusters. Only one sensor node is solely responsible to communicate to the base station in a cluster. This sensor node is called cluster head and the remaining sensor nodes in the cluster are called followers.

The followers collect data and send it to their corresponding cluster heads. The cluster heads aggregate its own data with the data received from its followers. Aggregated data is then sent to a sink to accomplish a specific goal. Cluster heads remain closer to their follower sensor nodes compared to the sink. It takes less energy to transmit data to the cluster head instead of the sink, which allows the sensor nodes to conserve more energy and live longer in WSN.

There are different clustering techniques already established for adhoc networks. However, those techniques cannot be directly used in WSN because of the fact that WSN imposes strict requirements on the energy efficiency than that adhoc networks do. As a result, many techniques have been proposed for clustering in WSN. Dynamic clustering techniques are more useful for WSN because of the dynamic variation in residual energies of the sensor nodes.

Some of the previous clustering techniques are: Distributed Clustering Algorithm (DCA) [3], Spanning Tree (or BFS Tree) based Clustering [4], Clustering with On-Demand Routing [5], Clustering based on Degree or Lowest Identifier Heuristics [6], Distributed and Energy-Efficient Clustering [7], Adaptive Power-Aware Clustering [8], Power-Efficient Gathering in Sensor Information Systems (PEGASIS), Power Efficient and Adaptive Clustering Hierarchy (PEACH), Optimal Energy Aware Clustering Algorithm for Cluster establishment (ACE), Hybrid Energy-Efficient Distributed Clustering (HEED).

Lindsey et al [9] introduced a near optimal chain-based protocol. Here, each node communicates only with a close neighbor and takes turns transmitting to the base station, thus reducing the amount of energy spent per round. It assumes that all nodes have global knowledge of the network and employ the greedy algorithm. It maps the problem of having close neighbors for all nodes to the traveling salesman problem. PEGASIS is a greedy chain protocol that is near optimal for a data-gathering problem in sensor networks. Greedy approach considers the physical distance only, ignoring the capability of a prospective node on the chain. Hence, a node with a shorter distance but less residual energy may be chosen in the chain and may die quickly.

It was proposed in [10] that a routing algorithm, which could combine hierarchical and geographical routing might performed well in greedy environments. The process of packet forwarding from the source nodes in the target region to the base station consists of two phases- inter-cluster routing and intra-cluster routing. For inter-cluster routing, a greedy algorithm is adopted to forward packets from the cluster heads of the target regions to the base station. For intracluster routing, a simple flooding is used to flood the packet inside the cluster when the number of intra-cluster nodes are less than a predetermined threshold. Otherwise, the recursive geographic forwarding approach is used to disseminate the packet inside target cluster, that is, the cluster head divides the target cluster into some sub regions, creates the same number of new copies of the query packet, and then disseminates these copies to a central node in each sub region.

PEACH [11] was a cluster formation technique based on overheard information from the sensor nodes. According to this approach, if a cluster head node becomes an intermediate node of a transmission, it first sets the sink node as its next hop. Then it sets a timer to receive and aggregate multiple packets from the nodes in the cluster set for a pre-specified time. It checks whether the distance between this node and the original destination node is shorter than that of between this node and already selected next hop node. If the distance is shorter, this node joins to the cluster of the original destination node and the next hop of this node is changed to the original destination node. PEACH is an adaptive clustering approach for multi-hop inter-cluster communication. However, it suffers from almost the same limitations of PEGASIS due to the choice of physical propinquity.

Optimal energy aware clustering [12] solves the balanced k-clustering problem optimally, where k signifies the number of master nodes that can be in the network. The algorithm is based on the minimum weight matching. It optimizes the sum of spatial distances between the member sensor nodes and the master nodes in the whole network. It effectively distributes the network load on all the masters and reduces the communication overhead and the energy dissipation. However, this research work does not consider of residual energy level while choosing a node as the master.

ACE [13] is a distributed clustering algorithm, which establishes clusters into two phases spawning and migration. There are several iterations in each phase and the gap between two successive iterations follows uniform distribution. During the spawning phase, new clusters are formed in a self-elective manner. When a node decides to become a cluster head, it will broadcast a message to its neighbors to become its followers. During migration phase, existing clusters are maintained and rearranged, if required. Migration of an existing cluster is controlled by the cluster head. Each cluster head will periodically poll all of its followers to determine which could be the best candidate to elect as a new leader for the cluster. Current cluster head will promote the best candidate as the new cluster head and abdicate itself from its position. ACE results in uniform cluster formation with a packing efficiency close to hexagonal close packing. However, ACE does not consider the residual energy of the nodes while selecting cluster heads.

Distributed algorithms called HEED [14] incorporates the residual energy of sensor nodes, which results in the formation of clusters by uniformly distributing the cluster heads across the network. It periodically selects cluster heads according to a hybrid parameter, which consists of a primary parameter, the residual energy of a node, and a secondary parameter, such as propinquity of a node to its neighbors or node degree. HEED converges in 0 (1) iterations using low messaging overhead and achieves uniform cluster head distribution across the network. However, it chooses the initial percentage of cluster heads randomly. This random choice remains as a severe limitation of this algorithm.

A number of research attempts to improve network stability period by various techniques like routing, scheduling, aggregation etc. However, in this paper we attempt to improve the network stability period using clustering as it can serve as a better platform for upper layer functionality such as broadcasting, aggregation etc. Our approach ESRPSDC exploits the underlying method of Energy-Efficient Level Based Clustering Routing Protocol [15]. In our proposal, we have incorporated the security methods as specified in [16].

## III. Model of Hybrid SNR based Dynamic Clustering

Grouping of sensor nodes into clusters has been widely pursued by many of the research communities in order to obtain network lifetime. Generally, the clustering methods can be categorized into static and dynamic clustering. The static clustering aims at minimizing the total energy spent during the formation of the clusters for a set of networks [17].

In this paper, we assume a sensor network model with following properties:

- All the sensor nodes are heterogeneous with limited supply of energy.
- Each node senses the data and transfers the information to the CH.
- The base station is located at a distance away from the base station and it is static.
- Each node has a fixed number of transmission power levels.
- Nodes are not equipped with Global Positioning System (GPS) unit.

The proposed system follows SNR based dynamic clustering and its process is divided in to five different phases namely Initialization, Energy based CH selection, SNR based CH selection by NCH nodes, Data forwarding through inter cluster routing, and Identifying the intruder.

### a. Initialization

As shown fig 1, after the deployment of the nodes, the base station broadcasts a request (REQ) message to every node .When the nodes have received the REQ, they are equally divided in to clusters depending on the number of nodes and its sensing range. Each cluster frames its own cluster ID and the cluster table (CT) as shown in Table 1.

Table 1. Initial Cluster table

| Cluster ID | No. of active nodes | No of sleep nodes | CH with its energy (Joules) | Next CH |
|---|---|---|---|---|
| 1 | 8 | 3 | Null | Null |
| 2 | 6 | 4 | Null | Null |

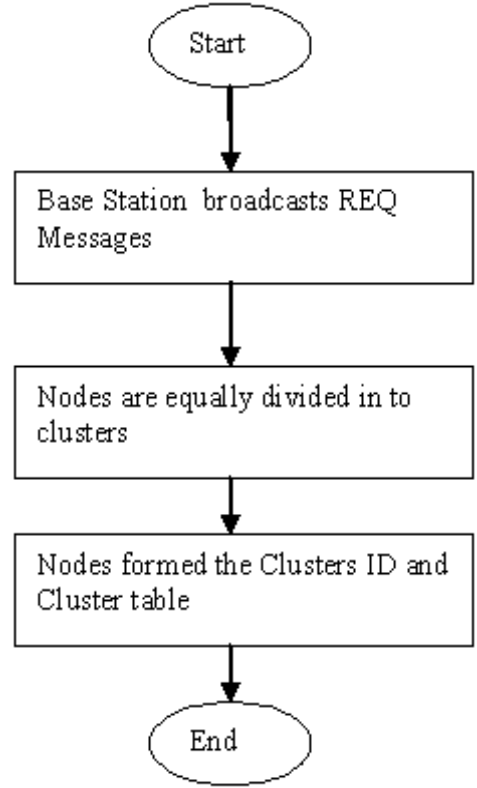

Fig 1. Initialization phase

The cluster table maintains cluster head node number alone with its energy. The nodes, which are alive, will be been considered as active nodes and which are turned off will be considered as sleep nodes. Initially during the creation of cluster table, the CH node number and its energy will be null.

On the initial deployment, the base station (BS) transmits a level-1 signal with minimum power level. All nodes, which hear this message, set their level as 1. After that, the base station increases its signal power to attain the next level and transmit a level-2 signal. All the nodes that receive the massage but do not set the previous level set their level as 2.

This procedure continuous until the base station transmits corresponding massages to all levels. The total number of messages of levels is equivalent to the number of distinct transmit signal at which the BS can sends. BS broadcast a hello massage, which contains the information of upper limit and lower limit of each level.

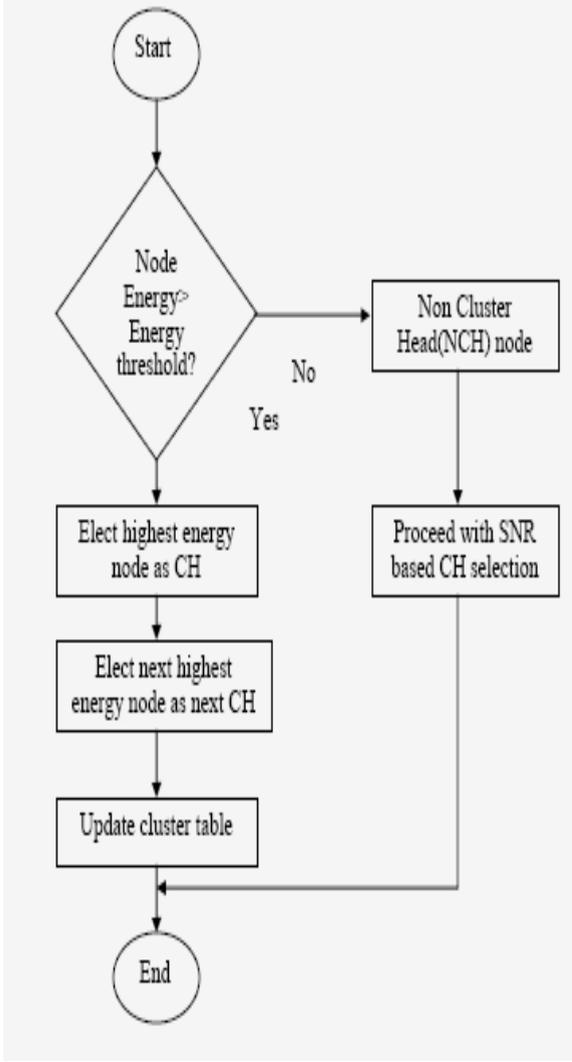

Fig 2. Energy based CH selection

### b. Energy based CH selection

Every Cluster group can elect its own cluster head based on its energy. Among, all the nodes in the cluster, the node, which is having the highest energy, have been chosen as CH [18]. The next highest energy node is chosen as next CH, so that during the next iteration if the CH losses its energy the next CH becomes the Current CH. The flow chart is shown in Fig.2. The threshold defined in equation 1.

$$T_i(n) = \frac{(P \times C)(U_i - d(n, BS))}{(1 - P)(r \bmod 1/P)(U_i - L_i)} \left[\frac{E_{cur}(n)}{E_{min}(n)}\right] K \quad \cdot (1)$$

where **P** is the desired percentage of the cluster heads. **r** is the current round. **Z** is the set of nodes, which have not been CHs in the last **1/P** rounds.

**C** is the constant factor between zero and one. Ui is the upper limit of level-i. 'Li' is the lower limit of level-i.

**d(n, BS)** will be the distance between node n and base station.

$E_{cur\,(n)}$ is defined as the current energy of node n.

$E_{min\,(n)}$ will be the initial energy of node **n** and the value of **K** will be between zero to three.

The updated cluster table is shown in table 2.

Table 2. Updated Cluster table

| Cluster ID | No. of active nodes | No of sleep nodes | CH with its energy (Joules) | Next CH with its energy (Joules) |
|---|---|---|---|---|
| 1 | 8 | 3 | N6/5 | N8/4.5 |
| 2 | 6 | 4 | N4/6.5 | N5/6 |

### c. SNR based CH selection by NCH nodes

In many cases, those nodes, which are distributed in sparse regions [19] or at the edge of a network, could not directly communicate with cluster heads due to limitation on their radio ranges. There are tradeoffs among connectivity, energy usage, and communication latency. In our work, communication between a cluster head and a node beyond the radio range of the cluster head has been achieved through intermediate nodes (1-hop member nodes) which provide relaying service based on their SNR value as shown in Fig.3. If normal node will receives a cluster head state message from the CH node and not belonging to any other cluster than it will send a confirm message to CH node. Now the normal node becomes a 1-hop node. It will create its own ID and send a state message to its neighbors within their region. If a Non Clustered Head (NCH) node receives a state message from a 1-hop member node, it will declare itself as a 2-hop member node. The two-hop member node also chooses its own ID, which is **m** byte random integer added at the end of the selected 1-hop member node's ID.

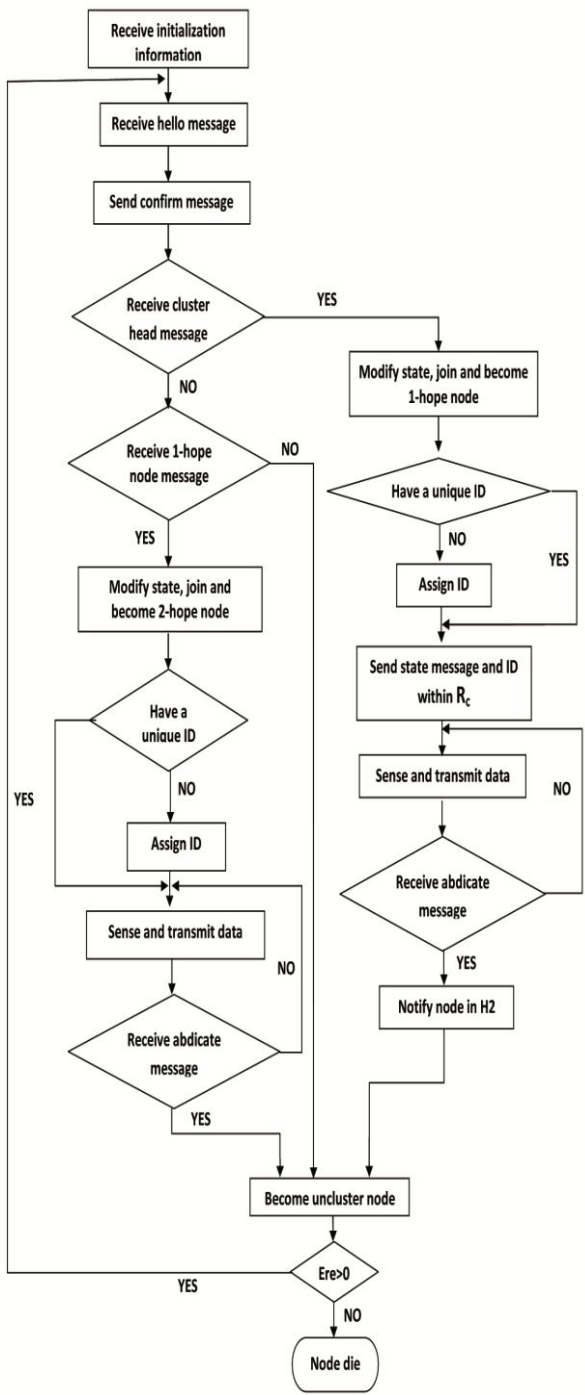

Fig 3. SNR based CH selection by NCH

It may rarely happen that two sensor nodes within a same cluster choose the same random number. This conflict can be solved through the cluster head by giving one of the nodes a different ID. Thus, at the end of this phase every node has its locally unique ID and knows which cluster it belongs. The abdicate message is sent by each cluster head to notify its member nodes of its unwillingness to serve as the cluster head in the next round, because of their lower energy levels

### d. Data forwarding through Inter cluster routing

After that, each cluster head creates [20] a TDMA schedule for its cluster members. This information is broadcasted back to the nodes in the cluster. Once the clusters are created and TDMA schedule is fixed, data transmission can begin. Each cluster member can be turned off until the node's allocated time. Each node sends data to its cluster heads with minimal transmission power. This power is estimated by received signal strength of the advertisement message, so that data transmission uses a minimal amount of energy. When all the data has been received from the cluster members, then cluster head node perform data aggregation function to compress the data into a single signal. After a certain time the next round, begin. After the cluster formation, the cluster heads broadcast the aggregate data to the next level. At the next level, the nodes aggregate their data and sends to their cluster heads. In this manner, the cluster heads at the last level transmit the final information to the BS.

### e. Identifying the intruder

Generally, the attacked area may contain many nodes and the intruder nodes are not necessarily located at the center of the area [21] in a multi-hop sensor network. Hence, it is necessary to further locate the exact intruders and isolate them from the network. This can be achieved through analyzing the routing pattern in the affected area. We now demonstrate a method for collecting the network flow information, which facilitates the routing pattern analysis. First, the Base Station (BS) sends a request message to the network. The message contains the IDs of the affected nodes, and is flooded hop by hop. For each node receiving the request, if its ID is there, it should respond to the BS with a message, which includes its own ID, the ID of the next-hop node, and the cost for routing, e.g, hop-count to the BS. Note that the next-hop and the cost could already be affected by the attack, hence, the response message should be transmitted along the reverse path in the flooding, which corresponds to the original route with no intruder.

The BS can then visualize the routing pattern by constructing a tree using the collected next hop information. Note that the area invaded by a sinkhole attack has a special routing pattern, where all network traffic flows toward the same destination, that is, the intruder Sink Hole (SH). As shown in Fig. 4, once the tree is constructed, the BS can easily identify the SH, which is exactly the root of the tree in this single malicious node case.

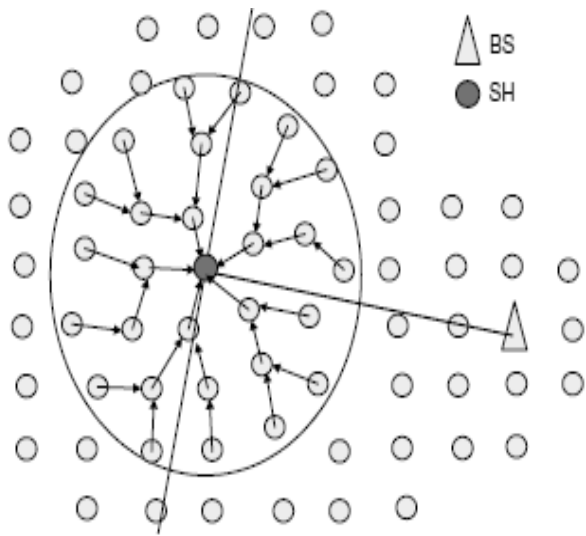

Fig 4. Pattern of the attacked area

## IV. Simulation Results

We used a simulation model based on GloMoSim-2.03 [22] in our evaluation. Our performance evaluations are based on the simulations of 500 wireless sensor nodes that form a wireless sensor network over a rectangular (1000 X 1000 m) flat space. The MAC layer protocol used in the simulations was the Distributed Coordination Function (DCF) of IEEE 802.11. The performance setting parameters were given in Table 3.

We fix the distance between the sources and sink to be 350 meters. The other 498 nodes were deployed between the source-sink pair. The performance setting parameters were given in Table 3.

Table 3.Simulation Parameters

| Parameter | Value |
| --- | --- |
| Area of sensing field | 1000 *1000 m |
| Number of sensor nodes | 1000 |
| Simulation Time | 600 s |
| Frequency | 2.4 GHz |
| Bandwidth | 2Mbps |
| Traffic Type | Variable Bit rate (VBR) |
| Payload Size | 30 to 70 Bytes |
| Number of Loads | 200 Packets |
| Number of Nodes | 500 nodes |
| Propagation Limit (dbm) | -111.0 |
| Path loss model | Two ray model |
| Location of the BS | (50, 75) |
| Number of clusters | 20 |
| Initial energy of nodes | 0.5J |
| Antenna Type | Omni directional |
| Channel Bandwidth | 20Kbps |
| MAC layer protocol | IEEE 802.11 |

We make the following assumptions about the WSN and the malicious node:

- WSN nodes are deployed uniformly at random in a planar square region. All nodes have the same wireless communication range, following the unit disk model.
- All nodes were implemented with ESRPSDC routing algorithm, and have loosely synchronized clocks.
- Every node sends one data packet at a random time during a specified send interval *S*. The payload of each packet indicates the originator of the data. No encryption mechanism has been deployed within the network.
- A single malicious node is present when the WSN is first deployed. The malicious node may be a compromised node or an implanted node. It has the same basic capabilities as legitimate sensor nodes.
- The malicious node participates in the network activities, but may provide false information in its link quality advertisements. The malicious node may also drop, modify, or divert the traffic that traverses it.

We compared our ESRPSDC with LEACH and PEGASIS, based on the following three parameters:

- **Packet Delivery Ratio:** It is the ratio of the successfully delivered data packets to the destinations to those generated by the CBR sources.

$$PDR = N_r / N_t \quad (2)$$

Where $N_r$ is the number of data packets successfully received and $N_t$ is the number of data packets transmitted.

- **End-to-End Delay** (Seconds): It indicates the time taken for the message to reach from source to destination.

- **Energy Consumption in milli Watt Hour (mWH).**

**Investigations-I :**

Investigations – I focus attention on comparing the packet delivery ratio when the load and Network size were varied in the presence of 30 % of malicious nodes.

The observations are as follows:

In general, packet delivery ratio decreases as the number of load and Network size were increased as shown in the Fig.5 and Fig.6. On an average, packet delivery ratio drops from 70.41% to 20.18% for ESRPSDC. For LEACH, packet delivery ratio has a steep fall from 32.07 % to 6.08% and in the case of PEGASIS, packet delivery ratio drops from 30.37% to 3.03%. Clearly, ESRPSDC delivers more packets than LEACH and PEGASIS.

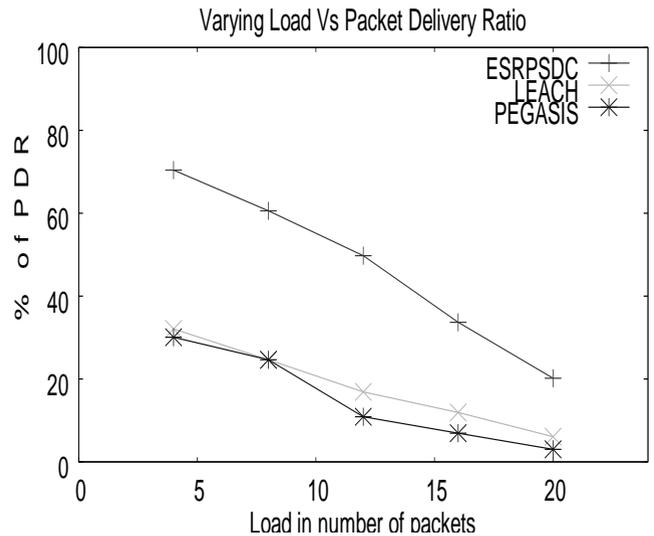

Fig 5. Comparison of **Load Vs %PDR** between ESRPSDC, LEACH and PEGASIS

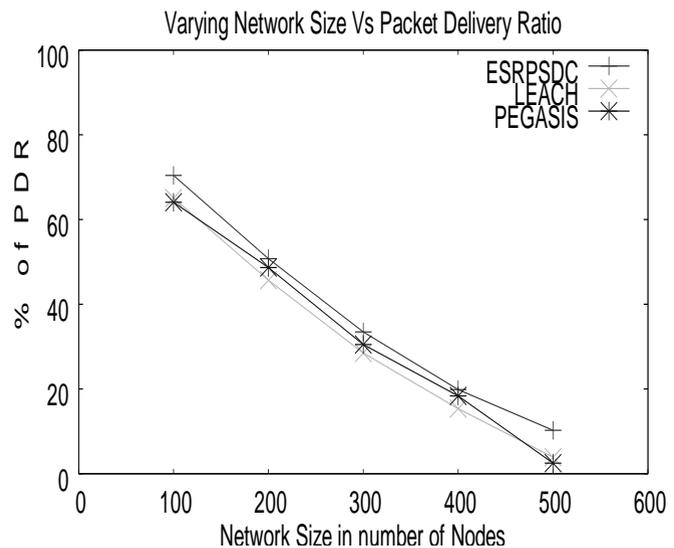

Fig 6. Comparison of **Network Size Vs %PDR** between ESRPSDC, LEACH and PEGASIS

**Investigations-II:**

Investigations – II focus attention on comparing the End to end delay in seconds, when the load and Network size were varied in the presence of 30 % of malicious nodes.

The observations are as follows:

As shown in Fig.7 and Fig.8, the End-to-End delay increases as the number of load and Network size were increased. The average increase in the case of ESRPSDC is from 5 Seconds to 32 Seconds. For LEACH, the delay increases form 3 seconds to 17 seconds and for PEGASIS the increase is from 3 seconds to 23 seconds.

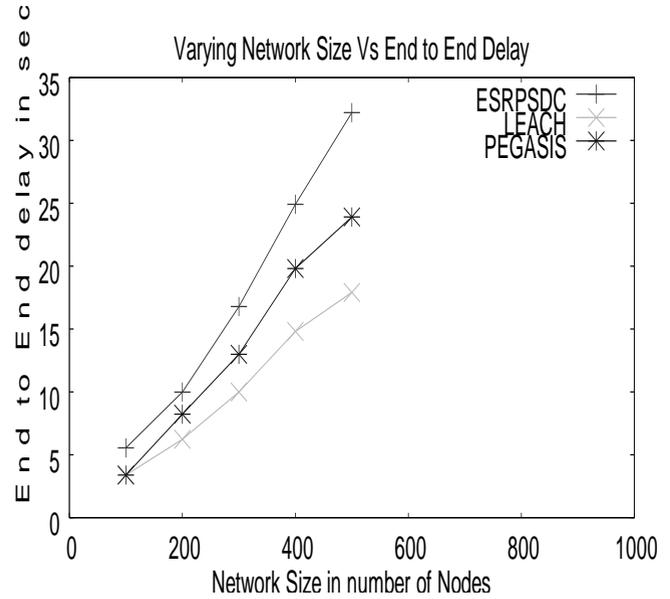

Fig 8. Comparison of Network **Size Vs End to End Delay** between ESRPSDC, LEACH and PEGASIS

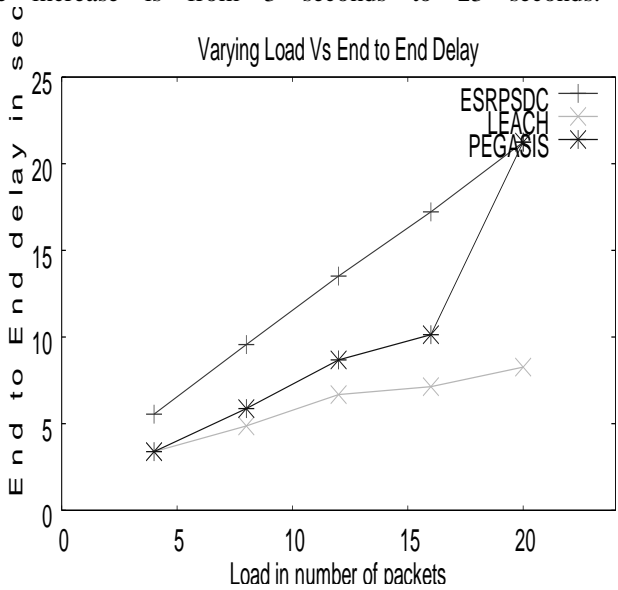

Fig 7. Comparison of **Load Vs End to End Delay** between ESRPSDC, LEACH and PEGASIS

The adverse increase in End-to-End delay could be observed in Fig.8 as compared to Fig.7, when the network size increases. The increased delay in ESRPSDC can be attributed to the increased level security mechanisms of it. LEACH and PEGASIS were performed well, when there was minimum number of malicious nodes. However, as shown in Fig.9, ESRPSDC performed better compared to LEACH and PEGASIS when the percentage of malicious nodes increases. This proves the efficiency of ESRPSDC security mechanism, since most of the malicious nodes were identified and isolated before the actual data transmission. During the data transmission, they were trapped and misdirected

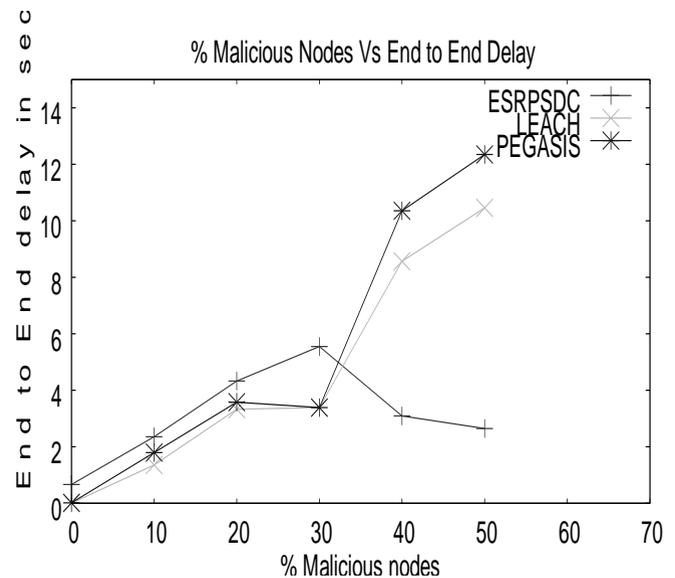

Fig 9. Comparison of %**Malicious nodes Vs End to End delay** between ESRPSDC, LEACH and PEGASIS

**Investigations-III :**

Investigations – III focus attention on comparing the average power consumption in mWH, when the Network sizes have been varied in the presence of 30 % of malicious nodes. The power consumption reduces in all three protocols, which shows their clustering strength, but nearly 50 % more reductions could be observed in

ESRPSDC compared to LEACH and PEGASIS as shown in Fig.10.

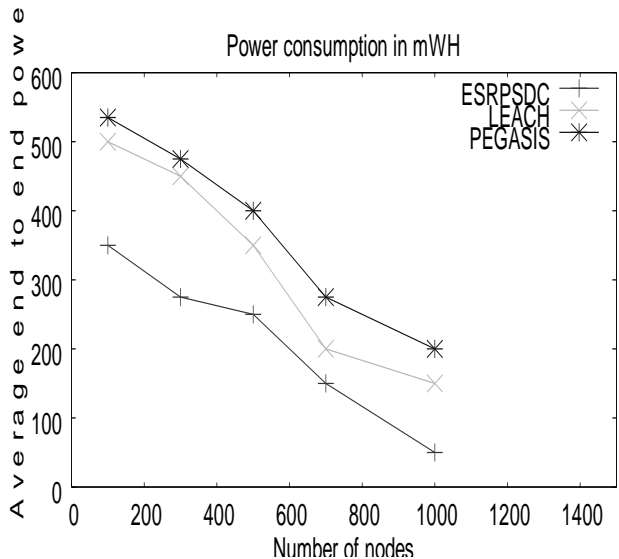

Fig 10. Comparison of Power Consumption in mWH between ESRPSDC, LEACH and PEGASIS

## V. Conclusions

The energy efficiency of a candidate route is critically dependent on the packet error rate of the underlying links, since they directly affect the energy wasted in retransmissions. Analysis of the interplay between error rates, number of hops, and transmission power levels reveals several key results. It has been shown in [23] that for reliable energy-efficient communication, the routing algorithm must consider both the distance and quality (e.g., in terms of the link error rate) of each link. Thus, the cost of choosing a particular link should be the overall transmission energy (including possible retransmissions) needed to ensure eventual error-free delivery, and not just basic transmission power. This is particularly important in practical multi hop wireless environments, where packet loss rates could be high [24].

In this paper, routing protocols for energy efficient data collection through SNR based dynamic clustering have been proposed. The network model based on power levels have been developed along with the mathematical formulae for choosing the cluster head. The developed model was simulated using GloMoSim. We have studied in detail about the simulation results of energy consumption of cluster heads, percentage packet delivery ratio and end-to-end delay.

Our future research might have focused about the optimization of our algorithm in order to effectively consume the energy of all nodes and improve the network lifetime. We shall extend our algorithm to heterogeneous WSNs.

The process of isolating the intruder or the compromised node could increase the number of hop count, which would further increases the delay in data delivery. Hence, node replacements strategies have to be analyzed carefully. In addition, we need to calculate the amount of overhead involved in our proposed scheme.